\documentclass[aps,showkeys,showpacs,superscriptaddress,a4paper,10pt]{revtex4-2}

\usepackage{amssymb}
\usepackage{amsmath}
\usepackage[dvips]{graphicx}
\usepackage[T2A]{fontenc}
\usepackage[dvips]{color}
\usepackage[dvipsnames]{xcolor}
\usepackage[breaklinks=true,colorlinks=true,linkcolor=blue,urlcolor=blue,citecolor=blue]{hyperref}

\begin{document}

\title[Sixth order modification of the Cahn-Hilliard equation]{Sixth order modification of the convective-viscous Cahn-Hilliard equation}

\author{P.~O.~Mchedlov-Petrosyan}
\author{L.~N.~Davydov}
\email[Corresponding author: ]{ldavydov@kipt.kharkov.ua}

\affiliation{National Science Center "Kharkiv Institute of Physics and Technology",   \\
1, Akademichna St., Kharkiv, 61108, Ukraine}

\author{O.~A.~Osmaev}
\affiliation{National Science Center "Kharkiv Institute of Physics and Technology",   \\
1, Akademichna St., Kharkiv, 61108, Ukraine}
\affiliation{Ukrainian State University of Railway Transport, 7, Feijerbakh Sq., Kharkiv, 61001, Ukraine}

\begin{abstract}
We consider the sixth-order convective-viscous Cahn-Hilliard equation, different from the standard fourth-order Cahn-Hilliard equation due to the modified expression for the thermodynamic potential. In such modified thermodynamic potential the coefficient at the square gradient term is order-parameter-dependent. It also contains the square of the Laplacian. This results in a sixth-order differential equation and additional nonlinear terms in the equation. We obtained several exact static- and traveling wave solutions and studied the dependence of solutions on the parameters of the system.
\end{abstract}


\maketitle

\renewcommand{\theequation}{\arabic{section}.\arabic{equation}}
\section{Introduction}\label{s1}

The present work is devoted to the study of the modified Cahn-Hilliard equation for the evolution of an order parameter $w$,
\begin{equation} \label{1.1} \frac{\partial w}{\partial t'} -2\bar{\alpha }w\frac{\partial w}{\partial x'} =\frac{\partial }{\partial x'} \left[M\frac{\partial }{\partial x'} \left(\bar{\mu }+\bar{\eta }_{1} \frac{\partial w}{\partial t'} -\bar{\eta }_{2} \frac{\partial ^{2} }{\partial x'^{2} } \frac{\partial w}{\partial t'} \right)\right] ,\end{equation}
\begin{equation} \label{1.2} \bar{\mu }=\left[\frac{1}{2} \frac{dg}{dw} \left(\frac{\partial w}{\partial x'} \right)^{2} -\frac{\partial }{\partial x'} \left(g\left(w\right)\frac{\partial w}{\partial x'} \right)\right]+\bar{\varepsilon }_{2} \frac{\partial ^{4} w}{\partial x'^{4} } +F\left(w\right) . \end{equation}

Here $\bar{\mu }$ is the modified chemical potential, corresponding to the double-well thermodynamic potential, which contains both $g\left(w\right)\left(\frac{\partial w}{\partial x'} \right)^{2} $ and $\left(\frac{\partial ^{2} w}{\partial x'^{2} } \right)^{2} $ terms. Here $g\left(w\right)=\bar{\theta }w^{2} +\bar{\varepsilon }_{1} $ is a quadratic polynomial and $F\left(w\right)$ a cubic polynomial. To make clear the meaning of the above modification, we need to give some insight into the history and existing modifications of this equation. The Cahn-Hilliard equation \cite{1,2,3,4} is now a well-established model in the theory of phase transitions as well as in several other fields. The basic underlying idea of this model is that for inhomogeneous system, e.g. system undergoing a phase transition, the thermodynamic potential (e.g. free energy) should depend not only on the order parameter $w$, but on its gradient as well. The idea of such dependence was introduced already by Van der Waals \cite{5} in his theory of capillarity. For an inhomogeneous system the local chemical potential $\mu $ is defined as the variational derivative of the thermodynamic potential functional. If the thermodynamic potential is the simplest symmetric -- quadratic -- function of the gradient this leads to the local chemical potential $\mu $ which depends on the Laplacian, or for the one-dimensional case -- on the second order derivative of the order parameter. The diffusional flux $J$ is proportional to the gradient of the chemical potential $\nabla \mu $; the coefficient of proportionality is called the mobility $M$ \cite{6}. With such an expression for the flux the diffusion equation instead of the usual second order equation becomes a forth-order PDE for the order parameter $w$ (herein our notations differ from the notations in the original papers):
\begin{equation} \label{1.3)} \frac{\partial w}{\partial t'} =\nabla \left[M\nabla \mu \right] , \end{equation}
\begin{equation} \label{1.4)} \mu =-\bar{\varepsilon }^{2} \Delta w+F\left(w\right) .\end{equation}

Here $M$ is the mobility, $\bar{\varepsilon }$ is usually presumed to be proportional to the capillarity length, and $F\left(w\right)=\frac{d\Psi \left(w\right)}{dw} $, where $\Psi \left(w\right)$ is the homogeneous part of the thermodynamic potential. Originally $F\left(w\right)$ was taken in the form of a cubic polynomial (corresponding to the fourth-order polynomial for the homogeneous part of the thermodynamic potential). The classic Cahn-Hilliard equation was introduced as early as in 1958 \cite{1,2}; the stationary solutions were considered, the linearized version was treated and the corresponding instability of homogeneous state identified. However, intensive study of the fully nonlinear form of this equation started essentially later \cite{7}. Now an impressive amount of work is done on the nonlinear Cahn-Hilliard equation, as well as on its numerous modifications, see \cite{3,4}. An important modification was done by Novick-Cohen \cite{8}. Taking into account the dissipation effects which were neglected in the derivation of the classic Cahn-Hilliard equation, she introduced the \textbf{\textit{viscous}} Cahn-Hilliard (VCH) equation
\begin{equation} \label{1.5} \frac{\partial w}{\partial t'} =\nabla \left[M\nabla \left(\mu +\bar{\eta }\frac{\partial w}{\partial t'} \right)\right] ,\end{equation}
where the coefficient $\bar{\eta }$ is called the viscosity. It was also noticed that the VCH equation could be derived as a certain limit of the classic Phase-Field model \cite{9}. Later several authors considered the nonlinear \textbf{\textit{convective}} Cahn-Hilliard equation (CCH) in one space dimension \cite{10,11,12},
\begin{equation} \label{1.6} \frac{\partial w}{\partial t'} -\bar{\alpha }w\frac{\partial w}{\partial x'} =\frac{\partial }{\partial x'} \left(\frac{\partial \mu }{\partial x'} \right). \end{equation}
Leung \cite{10} proposed this equation as a continual description of lattice gas phase separation under the influence of an external field. Similarly, Emmott and Bray \cite{12} proposed this equation as a model for the spinodal decomposition of a binary alloy in an external field \textit{E}. Witelski \cite{11} introduced the equation \eqref{1.6} as a generalization of the classic Cahn--Hilliard equation or as a generalization of the Kuramoto--Sivashinsky equation \cite{13,14} by including a nonlinear diffusion term. In \cite{10,11,12} and \cite{15,16} several approximate solutions and only two exact static kink and anti-kink solutions were obtained. The `coarsening' of domains separated by kinks and anti-kinks was also discussed. Also the convective Cahn-Hilliard equation with cubic nonlinearity in the convective term was introduced in \cite{17,18}. However, in the present work we will consider only the Burgers-type nonlinearity, as in \eqref{1.6}. To study the joint effects of nonlinear convection and viscosity, Witelski \cite{19} introduced the convective-viscous-Cahn--Hilliard equation (CVCHE) with a general symmetric double-well homogeneous part $\Psi \left(w\right)$ of the potential:
\begin{equation} \label{1.7} \frac{\partial w}{\partial t'} -\bar{\alpha }w\frac{\partial w}{\partial x'} =\frac{\partial }{\partial x'} \left[M\frac{\partial }{\partial x'} \left(\mu +\bar{\eta }\frac{\partial w}{\partial t'} \right)\right], \end{equation}
\begin{equation} \label{1.8)} \mu =-\bar{\varepsilon }^{2} \frac{\partial ^{2} w}{\partial x'^{2} } +\frac{d\Psi \left(w\right)}{dw} . \end{equation}
With a constraint imposed on the nonlinearity and the viscosity, the approximate traveling-wave solutions were obtained. In \cite{20} for equation \eqref{1.7} with a polynomial potential, and the balance between the applied field and the viscosity, several exact single- and two-wave solutions were obtained.

Another line of modifications of the nonlinear Cahn-Hilliard equation which attracted much interest recently is the generalization of the form of the thermodynamic potential including higher-order polynomials $\Psi \left(w\right)$, non-constant coefficient $g\left(w\right)$ at $\left(\nabla w\right)^{2} $ and higher-order terms $\left(\Delta w\right)^{2} ,\left(\nabla w\right)^{4} $ etc.,  resulting in the higher-order differential operator in the equation. In \cite{21} Novick-Cohen noticed, that if the system as a whole is essentially inhomogeneous, the coefficient $g$ should be nonconstant. Such a generalization of the viscous CH equation was studied numerically in \cite{22}. In the modeling of amphiphilic systems the free energy functionals with a sixth-power polynomial, nonconstant $g\left(w\right)$, and $\left(\Delta w\right)^{2} $-term where used systematically by Gomper and co-workers \cite{23,24,25,26}; this results in the sixth-order Cahn-Hilliard-type equation. This work was continued and elaborated by Pawlow and co-workers \cite{27,28,29}. They incorporated viscous effects associated with the rates of the order parameter and its spatial gradients. In our one-dimensional version \eqref{1.1}-\eqref{1.2} this corresponds to the terms $\bar{\eta }_{1} \frac{\partial w}{\partial t'} $ and $\bar{\eta }_{2} \frac{\partial ^{2} }{\partial x'^{2} } \frac{\partial w}{\partial t'} $. They considered initial/boundary value problem in 3D and obtained several rigorous mathematical results. The properties of sixth-order model without dissipation were discussed in \cite{30}. Similar model without dissipation, but with degenerate mobility was considered in \cite{31}.

Somewhat different form of the sixth order Cahn-Hilliard equation arises also in the atomistic models of phase transitions \cite{32,33,34}, based on the Phase Field Crystal approach \cite{35,36}. The basic idea of this approach is to use the simplest form of the free-energy functional, which is minimized by a spatial periodic pattern of the order parameter. In such a functional the coefficient at $\left(\nabla w\right)^{2} $ should be negative; it contains also the higher-order term $\sim \left(\nabla ^{2} w\right)^{2} $. Then again in the evolution equation for the order parameter the sixth-order derivative emerges.

Another reason to consider sixth- and higher order modifications of the Cahn-Hilliard equation is the regularization of strongly anisotropic problems \cite{36}. In this connection a mathematical analysis of arbitrary high order Cahn-Hilliard equations was performed \cite{37,38,39,40}.
An essentially different sixth-order modification was introduced by Savina et. al. \cite{41} as a model for the faceting of a growing crystalline surface; we will not consider this modification here.

To the best of our knowledge the exact solutions for any modifications of the sixth-order Cahn-Hilliard equation were not obtained. Our paper is organized as follows: in the Section \ref{s2} we outline the solution method and obtain an exact static wave solution for the sixth-order convective-viscous CH equation \eqref{1.1}-\eqref{1.2}. In Section \ref{s3} we obtain several traveling-wave solutions for this equation and study the dependence of the solution on the parameters of the system. In Section \ref{s4} we discuss our results.

\setcounter{equation}{0}
\section{Exact solution for the sixth order modification of the convective-viscous CH equation}\label{s2}

We assume that the cubic polynomial $F\left(w\right)$ has three real roots $\bar{a}_{1} <\bar{a}_{2} <\bar{a}_{3} $, corresponding to two minima and one maximum of the quartic polynomial in the thermodynamic potential. Then \eqref{1.2} becomes
\begin{eqnarray} \label{2.1}  \mu &&=\left[-\frac{1}{2} \frac{dg}{dw} \left(\frac{\partial w}{\partial x'} \right)^{2} -g\left(w\right)\frac{\partial ^{2} w}{\partial x'^{2} } \right]+\bar{\varepsilon }_{2} \frac{\partial ^{4} w}{\partial x'^{4} } \nonumber \\ &&+\rho \left(w-\bar{a}_{1} \right)\left(w-\bar{a}_{2} \right)\left(w-\bar{a}_{3} \right) . \end{eqnarray}

As it is usually presumed from symmetry considerations, we will use a quadratic dependence of the coefficient $g$ on the order parameter, $g=\theta w^{2} +\bar{\varepsilon }_{1} $.  Introducing the non-dimensional order parameter $u=\frac{w}{\bar{a}_{3} } $, non-dimensional coordinate $x=\frac{x'}{X} $ and non-dimensional time $t=\frac{t'}{T} $ (where it is convenient to define  $\, X=\sqrt{\bar{\varepsilon }_{1} } ;\, \, \, T=\frac{\bar{\varepsilon }_{1} }{M} $), we rewrite equations \eqref{1.1} and \eqref{2.1} in the non-dimensional form:
\begin{equation} \label{2.2} \frac{\partial u}{\partial t} -2\alpha u\frac{\partial u}{\partial x} =\frac{\partial ^{2} }{\partial x^{2} } \left(\mu +\eta _{1} \frac{\partial u}{\partial t} -\eta _{2} \frac{\partial ^{2} }{\partial x^{2} } \frac{\partial u}{\partial t} \right) ,\end{equation}
\begin{eqnarray} \label{2.3}  \mu &&=\left[-\theta u\left(\frac{\partial u}{\partial x} \right)^{2} -\left(\theta u^{2} +1\right)\frac{\partial ^{2} u}{\partial x^{2} } \right]+\varepsilon \frac{\partial ^{4} u}{\partial x^{4} }  \nonumber \\ &&+\rho \left(u-a_{1} \right)\left(u-a_{2} \right)\left(u-1\right) . \end{eqnarray}
We also introduced the following notations:
\begin{eqnarray} \label{2.4)}  \alpha &&=\bar{\alpha }\frac{\bar{a}_{3} T}{X} , \, \, \, \eta _{1} =\frac{\bar{\eta }_{1} }{T} , \, \, \, \eta _{2} =\frac{\bar{\eta }_{2} }{X^{2} T} , \, \, \, \theta =\frac{\bar{\theta }\bar{a}_{3}^{2} }{\bar{\varepsilon }_{1} } , \, \, \, \varepsilon =\frac{\bar{\varepsilon }_{2} }{X^{4} } =\frac{\bar{\varepsilon }_{2} }{\bar{\varepsilon }_{1}^{2} } , \, \, \,  \nonumber \\ \rho &&=\bar{\rho }\bar{a}_{3}^{2} , \, \, \, a_{i} =\frac{\bar{a}_{i} }{\bar{a}_{3} } , \, \, \, \mu =\frac{\bar{\mu }}{\bar{a}_{3} }  . \end{eqnarray}

Looking for a traveling wave solution, we introduce $z=x-vt$. Equations \eqref{2.2}, \eqref{2.3} take form
\begin{equation} \label{2.5} -\frac{d}{dz} \left(vu+\alpha u^{2} \right)=\frac{d^{2} }{dz^{2} } \left(\mu -\eta _{1} v\frac{du}{dz} +\eta _{2} v\frac{d^{3} u}{dz^{3} } \right) ,\end{equation}
\begin{eqnarray} \label{2.6}  \mu &&=\left[-\theta u\left(\frac{du}{dz} \right)^{2} -\left(\theta u^{2} +1\right)\frac{d^{2} u}{dz^{2} } \right]+\varepsilon \frac{d^{4} u}{dz^{4} }  \nonumber \\ &&+\rho \left(u-a_{1} \right)\left(u-a_{2} \right)\left(u-1\right) . \end{eqnarray}
Integrating \eqref{2.5} once we get
\begin{equation} \label{2.7} -\alpha \left(u^{2} +\frac{v}{\alpha } u+C_{1} \right)=\frac{d}{dz} \left(\mu -\eta _{1} v\frac{du}{dz} +\eta _{2} v\frac{d^{3} u}{dz^{3} } \right) .\end{equation}

We are looking for a solution which approaches values $u_{1} ,\, u_{2} $  at $\mp \infty $. For such a solution the simplest proper Ansatz is
\begin{eqnarray} \label{2.8}  \frac{du}{dz} &&=\kappa \left(u-u_{1} \right)\left(u-u_{2} \right)=\kappa \left[u^{2} -pu+q\right]  \nonumber \\ p&&=u_{1} +u_{2} ;\, \, \, q=u_{1} u_{2} .  \end{eqnarray}
At $\pm \infty $ the right-hand side of \eqref{2.7} is zero; i.e., the left-hand side should be zero too. This means, that the polynomial in the left-hand side of \eqref{2.7}  should coincide with the polynomial in the right-hand side of \eqref{2.8}. Setting $C_{1} =q$ and
\begin{equation} \label{2.9} v=-\alpha p=-\alpha \left(u_{1} +u_{2} \right) ,\end{equation}
we can rewrite the left-hand side of \eqref{2.7} as
\begin{equation} \label{2.10)}  -\alpha \left(u^{2} +\frac{v}{\alpha } u+C_{1} \right)=-\alpha \left[u^{2} -pu+q\right]=-\frac{\alpha }{\kappa } \frac{du}{dz} .  \end{equation}

Integrating \eqref{2.7} once more, we get
\begin{equation} \label{2.11} \frac{\alpha }{\kappa } u+\mu -\eta _{1} v\frac{du}{dz} +\eta _{2} v\frac{d^{3} u}{dz^{3} } =C_{2} . \end{equation}
It is convenient to introduce
\begin{equation} \label{2.12} \Phi \left(u\right)=-\left[\theta u\left(\frac{du}{dz} \right)^{2} +\left(\theta u^{2} +1\right)\frac{d^{2} u}{dz^{2} } \right]+\varepsilon \frac{d^{4} u}{dz^{4} } -\eta _{1} v\frac{du}{dz} +\eta _{2} v\frac{d^{3} u}{dz^{3} }  \end{equation}
and rewrite \eqref{2.11} as
\begin{equation} \label{2.13} \rho \left[u^{3} -\left(\sigma +1\right)u^{2} +\left(\zeta +\sigma +\frac{\alpha }{\kappa } \right)u-\zeta \right]+\Phi \left(u\right)=C_{2} . \end{equation}
Here we have denoted
\begin{equation} \label{2.14)} \sigma =a_{1} +a_{2} ;\, \, \, \zeta =a_{1} a_{2} . \end{equation}

Using the Ansatz \eqref{2.8} all higher-order derivatives in \eqref{2.12} could be easily calculated:
\begin{equation} \label{2.15} \frac{d^{2} u}{dz^{2} } =\kappa \left(2u-p\right)\frac{du}{dz} =\kappa ^{2} \left[2u^{3} -3pu^{2} +\left(2q+p^{2} \right)u-pq\right], \end{equation}
\begin{eqnarray} \label{2.16)}  \frac{d^{3} u}{dz^{3} } &&=\kappa ^{2} \left[6u^{2} -6pu+\left(2q+p^{2} \right)\right]\frac{du}{dz} \nonumber \\ &&=\kappa ^{3} \left[6u^{4} -12pu^{3} +\left(7p^{2} +8q\right)u^{2} -\left(p^{3} +8pq\right)u+q\left(2q+p^{2} \right)\right] , \end{eqnarray}
\begin{equation} \label{2.17)} \frac{d^{4} u}{dz^{4} } =\kappa ^{3} \left[24u^{3} -36pu^{2} +2\left(7p^{2} +8q\right)u-\left(p^{3} +8pq\right)\right]\frac{du}{dz} , \end{equation}
\begin{equation} \label{2.18)} \theta u\left(\frac{du}{dz} \right)^{2} =\theta \kappa \left(u^{3} -pu^{2} +qu\right)\frac{du}{dz} , \end{equation}
\begin{equation} \label{2.19} \left(\theta u^{2} +1\right)\frac{d^{2} u}{dz^{2} } =\kappa \left(2\theta u^{3} -\theta pu^{2} +2u-p\right)\frac{du}{dz} . \end{equation}

Evidently $\Phi \left(u\right)$ can be written as $\Phi =G\left(u\right)\frac{du}{dz} ;$ where the following expression for $G\left(u\right)$ can be obtained using \eqref{2.15}-\eqref{2.19}:
\begin{eqnarray} \label{2.20}  G\left(u\right)&&=3\kappa \left(8\varepsilon \kappa ^{2} -\theta \right)u^{3} +2\kappa \left(\theta p-18\varepsilon \kappa ^{2} p+3\eta _{2} v\kappa \right)u^{2}  \nonumber \\ &&+\kappa \left[-\theta q-2+2\varepsilon \kappa ^{2} \left(7p^{2} +8q\right)-6\eta _{2} v\kappa p\right]u \nonumber \\ &&+\left\{\kappa p-\varepsilon \kappa ^{3} \left(p^{3} +8pq\right)+v\left[\eta _{2} \kappa ^{2} \left(2q+p^{2} \right)-\eta _{1} \right]\right\}. \end{eqnarray}
Using \eqref{2.9} we eliminate $v$ from the latter expression
\begin{eqnarray} \label{2.21}  G\left(u\right)&&=3\kappa \left(8\varepsilon \kappa ^{2} -\theta \right)u^{3} +2\kappa p\left(\theta -18\varepsilon \kappa ^{2} -3\eta _{2} \alpha \kappa \right)u^{2} \nonumber  \\ &&+\kappa \left[-\theta q-2+2\varepsilon \kappa ^{2} \left(7p^{2} +8q\right)+6\eta _{2} \alpha \kappa p^{2} \right]u\nonumber  \\ &&+\left\{\kappa p-\varepsilon \kappa ^{3} \left(p^{3} +8pq\right)-\alpha p\left[\eta _{2} \kappa ^{2} \left(2q+p^{2} \right)-\eta _{1} \right]\right\} . \end{eqnarray}

The highest power term in equation \eqref{2.13} is, except for $\Phi \left(u\right)$, cubic. This means that for this equation to be satisfied identically for arbitrary $u$ the terms higher than linear in the right-hand side of \eqref{2.21} should be eliminated. Equating to zero the coefficients at $u^{3} $ and $u^{2} $ terms, we obtain two constraints on the parameters:
\begin{equation} \label{2.22}  \kappa ^{2} =\frac{\theta }{8\varepsilon } , \end{equation}
\begin{equation} \label{2.23} p\left(5\theta +12\eta _{2} \alpha \kappa \right)=0 . \end{equation}

There are two possibilities: either $p=0$, or
\begin{equation} \label{2.24}  12\alpha \kappa \eta _{2} =-5\theta . \end{equation}
Let us consider $p=0$ first. This means, evidently, $v=0$, $u_{1} =-u_{2}$, that is the solution is the symmetric static kink. The expression for $G\left(u\right)$ simplifies significantly:
\begin{equation} \label{2.25)} G\left(u\right)=\kappa \left(\theta q-2\right)u .\end{equation}
Correspondingly, for this case $\Phi \left(u\right)$ is
\begin{equation} \label{2.26)} \Phi \left(u\right)=\kappa \left(\theta q-2\right)u\frac{du}{dz} =\kappa ^{2} \left(\theta q-2\right)u\left(u^{2} +q\right). \end{equation}
Substitution of $\Phi \left(u\right)$ into \eqref{2.13} and setting $C_{2} =-\zeta \rho $ yields:
\begin{equation} \label{2.27)} \left[\rho +\kappa ^{2} \left(\theta q-2\right)\right]u^{2} -\rho \left(\sigma +1\right)u+\rho \left(\zeta +\sigma +\frac{\alpha }{\kappa } \right)+\kappa ^{2} \left(\theta q-2\right)q=0 . \end{equation}

The latter equation should be satisfied identically for arbitrary $u$; equating to zero the coefficients at all powers of $u$ we get three constraints: The first two constraints are:
\begin{equation} \label{2.28} \sigma +1=0 ,\end{equation}
\begin{equation} \label{2.29} -q=u_{2}^{2} =\frac{2}{\theta } \left(4\frac{\varepsilon \rho }{\theta } -1\right) . \end{equation}
For $u_{2} $ to be real the right-hand side of the latter equation should be positive; this constraint imposes the condition
\begin{equation} \label{2.30} 4\varepsilon \rho >\theta . \end{equation}
The third constraint is
\begin{equation} \label{2.31} \zeta -1+\frac{\alpha }{\kappa } -q=0 .\end{equation}

So, if the constraints \eqref{2.22}, \eqref{2.29}, \eqref{2.31} and \eqref{2.28}, i.e.,
\begin{equation} \label{2.32} a_{1} +a_{2} +1=0 \end{equation}
are satisfied, the solution of \eqref{2.8} is simultaneously a stationary solution of \eqref{2.2}, \eqref{2.3}. Integrating \eqref{2.8} and taking the position of the maximal value of the derivative $\left|\frac{du}{dx} \right|$ as $x=0$, we get
\begin{equation} \label{2.33)} u=-u_{2} \tanh \left(\kappa u_{2} x\right) .\end{equation}

This stationary solution exists even for $\alpha =0$; naturally, the viscosities $\eta _{1} ,\, \eta _{2} $ dropped out. The constraints \eqref{2.31} and \eqref{2.32} impose evident limitations on the stationary states of the potential and the applied field. E.g., if we in addition to \eqref{2.32} assume that the potential is symmetric, $a_{1} =-1$, or equivalently $a_{2} =0$,  then $\zeta =0$ and
\begin{equation} \label{2.34)} \, \frac{\alpha }{\kappa } +u_{2}^{2} =1 .\end{equation}
For $\alpha =0$ and a symmetric potential the values of $u$ at $\pm \infty $ coincide with the stationary values of the potential.
On the other hand, if $p\ne 0$ in \eqref{2.23}, a traveling wave solution exists. The parametric dependence of such a solution is more complicated and will be considered in the next Section.

\setcounter{equation}{0}
\section{Traveling wave solutions and their parametric dependence}\label{s3}

Returning to \eqref{2.23} and presuming $p\ne 0$, we rewrite the constraint \eqref{2.24}:
\begin{equation} \label{3.1} \alpha \kappa \eta _{2} =-\frac{5\theta }{12}.  \end{equation}
The viscosity $\eta _{2} $ is positive per definition, i.e., the $\kappa $ and $\alpha $ should have different signs. Using again \eqref{2.22}, we eliminate $\theta $, and using\eqref{2.24}, we eliminate $\alpha \kappa \eta _{2} $ from $G\left(u\right)$, see \eqref{2.21}; we also introduce $\lambda ={\eta _{1} \mathord{\left/ {\vphantom {\eta _{1}  \eta _{2} }} \right. \kern-\nulldelimiterspace} \eta _{2} } $:
\begin{eqnarray} \label{3.2)} G\left(u\right)&&=\kappa \left\{\left[8\varepsilon \kappa ^{2} q-2-6\varepsilon \kappa ^{2} p^{2} \right]u+\left[p\left(1-\frac{10}{3} \varepsilon \lambda \right)-\frac{4}{3} \varepsilon \kappa ^{2} qp+\frac{7}{3} \varepsilon \kappa ^{2} p^{3} \right]\right\}\nonumber  \\ &&=\kappa \left\{\left[\theta q-2-\frac{3}{4} \theta p^{2} \right]u+\left[p\left(1-\frac{10}{3} \varepsilon \lambda \right)-\frac{1}{6} \theta qp+\frac{7}{24} \theta p^{3} \right]\right\} . \end{eqnarray}

Using the latter expression we calculate $\Phi \left(u\right)$:
\begin{eqnarray} \label{3.3)}
\Phi \left(u\right)&&=\kappa ^{2} \left(\theta q-2-\frac{3}{4} \theta p^{2} \right)u^{3} \nonumber \\ &&+\kappa ^{2} \left\{\left[p\left(1-\frac{10}{3} \varepsilon \lambda \right)-\frac{1}{6} \theta qp+\frac{7}{24} \theta p^{3} \right]-p\left(\theta q-2-\frac{3}{4} \theta p^{2} \right)\right\}u^{2} \nonumber \\ &&+\kappa ^{2} \left\{-p\left[p\left(1-\frac{10}{3} \varepsilon \lambda \right)-\frac{1}{6} \theta qp+\frac{7}{24} \theta p^{3} \right]+\, q\left(\theta q-2-\frac{3}{4} \theta p^{2} \right)\right\}u \nonumber \\ &&+\kappa ^{2} q\left[p\left(1-\frac{10}{3} \varepsilon \lambda \right)-\frac{1}{6} \theta qp+\frac{7}{24} \theta p^{3} \right].
 \end{eqnarray}
Substitution of the latter expression into \eqref{2.13}, selecting the constant $C_{2} $ to cancel the $u$-independent terms and equating to zero the coefficients at all powers of $u$ yields three constraints on the parameters,
\begin{equation} \label{3.4} \frac{\rho }{\kappa ^{2} } +\theta q-2-\frac{3}{4} \theta p^{2} =0 , \end{equation}
\begin{equation} \label{3.5} -\rho \left(\sigma +1\right)+\left[p\left(1-\frac{10}{3} \varepsilon \lambda \right)-\frac{1}{6} \theta qp+\frac{7}{24} \theta p^{3} \right]\kappa ^{2} -p\kappa ^{2} \left(\theta q-2-\frac{3}{4} \theta p^{2} \right)=0 ,\end{equation}
\begin{equation} \label{3.6} \rho \left(\zeta +\sigma +\frac{\alpha }{\kappa } \right)-p\kappa ^{2} \left[p\left(1-\frac{10}{3} \varepsilon \lambda \right)-\frac{1}{6} \theta qp+\frac{7}{24} \theta p^{3} \right]+\, q\kappa ^{2} \left(\theta q-2-\frac{3}{4} \theta p^{2} \right)=0 . \end{equation}

If the constraints \eqref{2.9}, \eqref{2.22}, \eqref{2.24}, and \eqref{3.4}-\eqref{3.6} are satisfied, the solution of \eqref{2.8} is simultaneously a solution of \eqref{2.5}-\eqref{2.6}. Integrating \eqref{2.8}, we get now
\begin{equation} \label{3.7)} u=\frac{u_{2} +u_{1} }{2} -\frac{u_{2} -u_{1} }{2} \tanh \left(\frac{1}{2} \kappa \left(u_{2} -u_{1} \right)\left(x-vt\right)\right) .\end{equation}

There are six constraints and only four unknowns $\kappa ,\, u_{1} ,\, u_{2} ,$ and $v$. This means that for such a solution to exist two constraints should be imposed on the parameters of the system. One constraint is \eqref{3.1}; we use \eqref{3.4}-\eqref{3.5} to find $p,q$ (i.e., $u_{1} ,\, u_{2} $), so the second constraint on the system parameters is \eqref{3.6}.

The necessary condition for the existence of the stationary solution was the constraint \eqref{2.28}, i.e., the limitation imposed on the stationary states of the thermodynamic potential. It is natural to presume, that this limitation should persist for the traveling wave as well; then \eqref{3.5} becomes

\begin{equation} \label{3.8} \left[\left(1-\frac{10}{3} \varepsilon \lambda \right)-\frac{1}{6} \theta q+\frac{7}{24} \theta p^{2} -\left(\theta q-2-\frac{3}{4} \theta p^{2} \right)\right]p=0 .\end{equation}

As we already know, see Section \ref{s2}, $p=0$ corresponds to the static solution; so here we presume $p\ne 0$, and \eqref{3.8} becomes
\begin{equation} \label{3.9)} \left(1-\frac{10}{3} \varepsilon \lambda \right)-\frac{1}{6} \theta q+\frac{7}{24} \theta p^{2} -\left(\theta q-2-\frac{3}{4} \theta p^{2} \right)=0 .\end{equation}
Using \eqref{3.4} we rewrite the latter equation as
\begin{equation} \label{3.10}  6-20\varepsilon \lambda -\theta q+\frac{7}{4} \theta p^{2} +6\frac{\rho }{\kappa ^{2} } =0 . \end{equation}
Summing \eqref{3.4} and \eqref{3.10} we get an equation for $p$
\begin{equation} \label{3.11} p^{2} =\frac{4}{\theta } \left(5\varepsilon \lambda -14\frac{\rho \varepsilon }{\theta } -1\right). \end{equation}
Substituting \eqref{3.11} into \eqref{3.4} yields
\begin{equation} \label{3.12)} q=\frac{1}{\theta } \left(15\varepsilon \lambda -50\frac{\rho \varepsilon }{\theta } -1\right) .\end{equation}
The discriminant of the quadratic equation $y^{2} -py+q=0$ is
\begin{equation} \label{3.13)} p^{2} -4q=\frac{8\varepsilon }{\theta } \left(18\frac{\rho }{\theta } -5\lambda \right) .  \end{equation}
Both $p^{2} $ and the discriminant should be non-negative; these inequalities define upper and lower limits on $\lambda $
\begin{equation} \label{3.14} \frac{14}{5} \frac{\rho }{\theta } +\frac{1}{5\varepsilon } \le \lambda \le \frac{18}{5} \frac{\rho }{\theta } . \end{equation}

\begin{figure}
\includegraphics[width=6 cm]{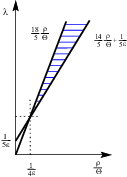}
\caption{ Illustration of the condition \eqref{3.14}: the allowed domain in the $\left(\frac{\rho }{\theta } ,\, \lambda \right)$ plane is hatched in blue. }\label{Fig:1}
\end{figure}

The condition \eqref{3.14} is illustrated in the Fig.\ref{Fig:1}; the allowed domain in the $\left(\frac{\rho }{\theta } ,\, \lambda \right)$ plane is hatched in blue. For \eqref{3.14} to be possible there should be$\, \frac{1}{4\varepsilon } <\frac{\rho }{\theta } ; $i.e., for the non-zero interval of allowed values of $\lambda $ to exist $\frac{\rho }{\theta } $ should be large enough. We remind that this is exactly the condition \eqref{2.30} for the stationary states of the static solution to be real. There are two solutions, corresponding to positive or negative $p$. From \eqref{2.9} $v=-\alpha p$; so if (for definiteness) $\alpha <0$, the anti-kink ($\kappa >0$) with positive $p$ moves in the positive direction, and the anti-kink with negative $p$ moves in the negative direction:
\begin{equation} \label{3.15} v=\pm 2\left|\alpha \right|\left[\frac{1}{\theta } \left(5\varepsilon \lambda -14\frac{\rho \varepsilon }{\theta } -1\right)\right]^{\frac{1}{2} } . \end{equation}
Both waves have the same amplitude $A=u_{2} -u_{1} $:
\begin{equation} \label{3.16)} A=2\left[\frac{2\varepsilon }{\theta } \left(18\frac{\rho }{\theta } -5\lambda \right)\right]^{\frac{1}{2} } . \end{equation}

Calculating the constraint \eqref{3.6} by using \eqref{3.4} and \eqref{3.5} we get
\begin{equation} \label{3.17} \zeta -1+\frac{\alpha }{\kappa } +p^{2} -q=0 .\end{equation}
Remarkably this constraint differs from the constraint \eqref{2.31} for the static solution by the presence of the $p^{2} $-term only.

\setcounter{equation}{0}
\section{Discussion}\label{s4}

Summing up, we have obtained exact static and traveling wave solutions for the modified Convective-Viscous Cahn-Hilliard equation. The modification results from the non-constant coefficient $g\left(w\right)$ at the $\left(\frac{\partial w}{\partial x} \right)^{2} $ term and the $\left(\frac{\partial ^{2} w}{\partial x^{2} } \right)^{2} $ term in the thermodynamic potential. This results in a sixth-order differential equation with additional nonlinear terms. While the form of solution is simple, and is the same as for the fourth-order convective-viscous Cahn-Hilliard equation \cite{19,20}, the nonlinear algebraic system, linking the parameters of solution to the parameters of the system, is more complicated. For exact static and traveling-wave solutions to exist additional constraints should be imposed on the parameters of the system. It was shown in \cite{20} that the necessary condition for the existence of the exact traveling wave solution for the fourth order convective-viscous CH equation is the balance between the forcing and the dissipation, i.e., the applied field $\alpha $ and the viscosity $\eta $. Remarkably, a similar condition exists in the present case, see \eqref{3.1}. Returning to the initial dimensional parameters we rewrite \eqref{3.1} as
\begin{equation} \label{4.1)} \frac{\bar{\eta }_{2} \bar{\alpha }}{\sqrt{2\bar{\theta }\bar{\varepsilon }_{2} } } =-\frac{5}{6} . \end{equation}

Here we presumed $\kappa $ to be positive (anti-kink wave), so $\bar{\alpha }$ should be negative. Interesting, while the numerator is, similar to \cite{20}, the product of coefficient of the convective term and the ``second viscosity'', the denominator is the geometric mean of the coefficients at the highest-order derivative terms. While for the existence of the exact traveling-wave solutions of the fourth-order convective-viscous Cahn-Hilliard equation only a single constraint on the system parameters was needed \cite{20}, for the sixth-order modification there are two constraints. The second constraint \eqref{3.17} connects the stationary states of the potential with the values of the solution at $\pm \infty $. For the static solution $p=0$, and Eq.~\eqref{3.17} coincides with the constraint \eqref{2.31}. For the existence of real $u_{1} ,\, u_{2} $ the ratio of the viscosities is limited from above and below, see \eqref{3.14} and Fig.1; these limits are determined by the parameters of the thermodynamic potential. So there is only a limited parametric interval, within which the balance of the ``driving force'' and the dissipation persists. Both for the stationary solution, see \eqref{2.30}, and the traveling wave the necessary condition is $\, 4\varepsilon \rho >\theta $. In the initial dimensional parameters this condition is
\begin{equation} \label{4.2)} 4\bar{\varepsilon }_{2} \bar{\rho }>\bar{\theta }\bar{\varepsilon }_{1}.  \end{equation}

The steepness of the front is determined by
\begin{equation} \label{4.3)} \kappa \left(u_{2} -u_{1} \right)=\left(18\frac{\rho }{\theta } -5\lambda \right)^{\frac{1}{2} } =\left[\bar{\varepsilon }_{1} \left(18\frac{\bar{\rho }}{\bar{\theta }} -5\frac{\bar{\eta }_{1} }{\bar{\eta }_{2} } \right)\right]^{\frac{1}{2} } . \end{equation}
So the steepness decreases with $\theta $ i.e., with the ``non-constancy'' of the coefficient at the $\left(\frac{\partial w}{\partial x} \right)^{2} $-term. As it was noticed by Novick-Cohen \cite{21}, the non-constant coefficient at $\left(\frac{\partial w}{\partial x} \right)^{2} $ term reflects the essential inhomogeneities of the system; i.e., the decreasing steepness of the front is due to these inhomogeneities. Expressed in the initial dimensional parameters the dimensional velocity is, see \eqref{3.15},
\begin{equation} \label{4.4)} \, \bar{v}=v\frac{X}{T} =\pm \frac{5\bar{\varepsilon }_{2} }{3\bar{\eta }_{2} } \left[2\left(5\frac{\bar{\eta }_{1} }{\bar{\eta }_{2} } -14\frac{\bar{\rho }\, }{\bar{\theta }} -\frac{\bar{\varepsilon }_{1} }{\bar{\varepsilon }_{2} } \right)\right]^{\frac{1}{2} }.  \end{equation}

Naturally, the velocity is determined by the competition of the driving force (the thermodynamic potential) and dissipation (the viscosity). Considering the absolute value of $\bar{v}$ as a function of two ratios $\gamma ={\bar{\varepsilon }_{2} \mathord{\left/ {\vphantom {\bar{\varepsilon }_{2}  \bar{\varepsilon }_{1} }} \right. \kern-\nulldelimiterspace} \bar{\varepsilon }_{1} } ;\, \, \xi ={\bar{\eta }_{2} \mathord{\left/ {\vphantom {\bar{\eta }_{2}  \bar{\eta }_{1} }} \right. \kern-\nulldelimiterspace} \bar{\eta }_{1} } $, while all other parameters kept fixed,
\begin{equation} \label{4.5)} \left|\bar{v}\right|=\frac{5\gamma }{3\xi } \frac{\bar{\varepsilon }_{1} }{\bar{\eta }_{1} } \left[2\left(\frac{5}{\xi } -14\frac{\bar{\rho }\, }{\bar{\theta }} -\frac{1}{\gamma } \right)\right]^{\frac{1}{2} } ,  \end{equation}
we see that with the relative increase of the coefficient at highest-order derivative term the velocity increases, and with the relative increase of the coefficient at the ``second viscosity'' term the velocity decreases. So the velocity of the phase transformation front is essentially influenced by the presence of the higher derivative terms, i.e., by the presumed larger inhomogeneity.

{\small \topsep 0.6ex

}

\end{document}